\documentclass[superscriptaddress]{revtex4}%
\usepackage{amsfonts}
\usepackage{amsmath}
\usepackage{amssymb}
\usepackage{graphicx}%
\providecommand{\U}[1]{\protect\rule{.1in}{.1in}}
\newtheorem{theorem}{Theorem}

\newtheorem{lemma}[theorem]{Lemma}

\begin{document}
\title{Analyzing the spin-bath model without simulations}
\author{Sebastian Fortin}
\affiliation{CONICET, IAFE (CONICET-UBA) and FCEN (UBA), Argentina}
\author{Olimpia Lombardi}
\affiliation{CONICET and FCEN (UBA), Argentina}
\keywords{Decoherence, discrete spectrum, closed systems, open systems, spin-bath model}
\pacs{03.65.Yz, 03.67.Bg, 03.67.Mn, 03.65.Db, 03.65.Ta, 03.65.Ud}

\begin{abstract}
On the basis of a lemma designed to decide whether a discrete system decoheres
or not with no need of computer simulations, in this paper we analyze the
well-known spin-bath model. The lemma allows us to predict the decoherence of
the system by analytical means.

\end{abstract}
\maketitle

\section{Introduction}

In previous papers we have developed a \textit{general theoretical framework
}for decoherence (\cite{CQG-General}, \cite{JPA-Sebi}, \cite{PLA},
\cite{Letter-1}), which can be applied to open and closed systems. The
conceptually relevant step in this framework is the selection of the relevant
observables, relative to which the question about decoherence is posed. Then,
decoherence can be explained in three steps:

\begin{enumerate}
\item \textbf{First step:} The space $\mathcal{O}_{R}$ of relevant observables
is defined.

\item \textbf{Second step:} The expectation value $\langle O_{R}\rangle
_{\rho(t)}$, for any $O_{R}\in\mathcal{O}_{R}$, is obtained.

\item \textbf{Third step:} It is proved that $\langle O_{R}\rangle_{\rho
(t)}=\langle O_{R}\rangle_{\rho_{R}(t)}$ reaches a final equilibrium value:
\begin{equation}
\lim_{t\rightarrow\infty}\langle O_{R}\rangle_{\rho(t)}=\lim_{t\rightarrow
\infty}\langle O_{R}\rangle_{\rho_{R}(t)}=\langle O_{R}\rangle_{\rho_{\ast}%
}=\langle O_{R}\rangle_{\rho_{R\ast}}\text{\ \ \ \ \ \ \ \ \ }\forall O_{R}%
\in\mathcal{O}_{R}\label{INT-01}%
\end{equation}

\end{enumerate}

As it is explained in another paper of this issue (\cite{Discreto}), this
general framework strictly applies when the limit of eq. (\ref{INT-01})
exists, and this happens when the Riemann-Lebesgue theorem is valid. But the
Riemann-Lebesgue theorem strictly applies only in cases of continuous energy
spectrum. Nevertheless, it can be shown that the validity conditions of the
discrete analogue of the Riemann-Lebesgue theorem are expressed by the
following lemma:

\begin{lemma}
\textit{Let} $\left\{  x_{i}\right\}  $ \textit{be a set of points uniformly
distributed, and }$f(x_{i}):\mathbb{R}\rightarrow\mathbb{R}$ \textit{be a
discrete function defined over }$\left\{  x_{i}\right\}  $, such that:

\begin{itemize}
\item $i\in\left[  0,N\right]  $ \textit{and} $N\gg1$.

\item $\exists G\in\mathbb{N},\exists P\in\mathbb{N}$ such that $P\gg1$ and
$\left\{  x_{i}\right\}  =\bigcup_{k=1}^{G}\left\{  x_{\left(  k-1\right)
\left(  P+1\right)  +1},...,x_{k\left(  P+1\right)  }\right\}  =\bigcup
_{k=1}^{G}X_{k}$. In this case we will say that the set $\left\{
x_{i}\right\}  $ \textit{is} \textit{quasi-continuous of class 1.}

\item $\forall X_{k},$ $f(x_{r_{k}})\cong C_{k}$, with $x_{r_{k}}\in X_{k}$.
In this case we will say that $f(x_{i})\in\mathcal{L}_{1}$\ .
\end{itemize}
\end{lemma}

\textit{then,}%
\begin{equation}
\ \lim_{t\longrightarrow t_{P}/2}\sum_{i=0}^{N}\frac{1}{N}f\left(
x_{i}\right)  e^{ix_{i}t}\cong0\label{Lemma}%
\end{equation}
\textit{where }$t_{P}$\textit{ is the recurrence or Poincar\'{e}
time}.\bigskip

The general aim of this paper is to apply this lemma to the spin-bath model,
in order to show that this method allows us to predict decoherence by
analytical means. For this purpose, the paper is organized as follows. In
Section 2 we will explain how the general framework for decoherence applies to
open systems, in particular, to models traditionally treated by means of the
environment-induced decoherence (EID) approach (\cite{Zeh-1970}%
-\cite{Zurek-2003}). In Section 3 we will present the spin-bath model, showing
which the relevant observables are in this case and how the expectation values
have to be computed. Section 4 will be devoted to compare two methods for
analyzing the model: the standard method, based on computer simulations, and
the analytical method, based on our lemma. \ Finally, in Section 5 we will
draw our conclusions.

\section{EID from the general framework}

In the case of the EID approach, the three steps of the general framework for
decoherence are usually not explicit in the formalism. However, the theory can
be rephrased in such a way that it can be analyzed from that framework. In
this section we will undertake this task in order to apply our just introduced
lemma in a following section.\bigskip

1. \textbf{First step:} Let us consider a closed system $U$ that can be
decomposed into a proper system $S$ and its environment $E$. Let the Hilbert
space of $U$ be $\mathcal{H}=\mathcal{H}_{S}\otimes$ $\mathcal{H}_{E}$, where
$\mathcal{H}_{S}$ is the Hilbert space of $S$ and $\mathcal{H}_{E}$ the
Hilbert space of $E$. The corresponding von Neumann-Liouville space of $U$ is
$\mathcal{L}=\mathcal{H\otimes H=L}_{S}\otimes$ $\mathcal{L}_{E}$, where
$\mathcal{L}_{S}=\mathcal{H}_{S}\otimes$ $\mathcal{H}_{S}$ and $\mathcal{L}%
_{E}=\mathcal{H}_{E}\otimes$ $\mathcal{H}_{E}$. A generic observable belonging
to $\mathcal{L}$ reads
\begin{equation}
O=\sum_{I}O_{S}^{(I)}\otimes O_{E}^{(I)}\text{ }\in\mathcal{L}\text{, \quad
with }O_{S}^{(I)}\in\mathcal{L}_{S}\text{ and }O_{E}^{(I)}\in\mathcal{L}_{E}
\label{EID-01}%
\end{equation}
i.e. $O$ is an observable with coordinates $(O_{i\alpha j\beta}^{(I)}%
)=(\sum_{I}O_{ij}^{(I)}O_{\alpha\beta}^{(I)})$, where $i,j,...$ are the
indices corresponding to $\mathcal{H}_{S}$, and $\alpha,\beta,...$ are the
indices corresponding to $\mathcal{H}_{E}$. The relevant observables are those
having the following form:%
\begin{equation}
O_{R}=O_{S}\otimes I_{E}\in\mathcal{O}_{R}\text{, \quad with coordinates
}(O_{ij}\delta_{\alpha\beta}) \label{EID-02}%
\end{equation}
where $I_{E}$ is the identity operator in $\mathcal{L}_{E}$. Therefore,
$\mathcal{O}_{R}\subset\mathcal{L}$ is the subspace of the relevant
observables, in this EID case those essentially corresponding to the proper
system $S$.\bigskip

2. \textbf{Second step:} The expectation value of any observable $O_{R}\in$
$\mathcal{O}_{R}$ in the state $\rho$ of $U$ reads%
\begin{equation}
\langle O_{R}\rangle_{\rho}=Tr\,(\rho O_{R})=\sum_{ij\alpha\beta}\rho_{i\alpha
j\beta}^{\ast}\,O_{ij}\,\delta_{\alpha\beta}=\sum_{ij}O_{ij}\sum_{\alpha\beta
}\rho_{i\alpha j\beta}^{\ast}\,\delta_{\alpha\beta}=\sum_{ij}O_{ij}%
\sum_{\alpha}\rho_{i\alpha j\alpha}^{\ast} \label{EID-03}%
\end{equation}
The reduced density operator $\rho_{R}$ is defined by tracing over the
environmental degrees of freedom,%
\begin{equation}
\rho_{S}=Tr_{E}\,\rho\in\mathcal{L}_{S}^{\prime}\text{, \quad with coordinates
}\left(  \sum_{\alpha}\rho_{i\alpha j\alpha}\right)  =(\rho_{ij})
\label{EID-04}%
\end{equation}
where $\mathcal{L}_{S}^{\prime}$ is the dual space of $\mathcal{L}_{S}$.
Therefore, the expectation value $\langle O_{R}\rangle_{\rho(t)}$ can be
expressed as
\begin{equation}
\langle O_{R}\rangle_{\rho(t)}=Tr\,\left(  \rho(t)\,O_{R}\right)  =Tr\,\left(
\rho(t)(O_{S}\otimes I_{E})\right)  =Tr\left(  \rho_{S}(t)\,O_{S}\right)
=\langle O_{S}\rangle_{\rho_{S}(t)} \label{EID-05}%
\end{equation}
\medskip

3. \textbf{Third step:} The EID approach studies the time evolution of the
reduced density operator $\rho_{S}(t)$ governed by an effective master
equation. For many physical models where the space $\mathcal{O}_{R}$ has a
finite number of dimensions, this approach shows that, for $t\rightarrow
\infty$, $\rho_{S}(t)$ reaches an equilibrium state $\rho_{S\ast}$:%
\begin{equation}
\rho_{S}(t)\longrightarrow\rho_{S\ast} \label{EID-06}%
\end{equation}
Since $\rho_{S\ast}$ is obviously diagonal in its eigenbasis, the system $S$
decoheres in the eigenbasis of $\rho_{S\ast}$, which turns out to be the final
decoherence basis. But if we take into account the definition of $\rho_{S}$ as
a partial trace (see eq. (\ref{EID-04})), we can obtain the limit of the
expectation values of eq. (\ref{EID-05}) as%
\begin{equation}
\lim_{t\rightarrow\infty}\langle O_{S}\rangle_{\rho_{S}(t)}=\lim
_{t\rightarrow\infty}\langle O_{R}\rangle_{\rho(t)}=\langle O_{S}\rangle
_{\rho_{S_{\ast}}}=\langle O_{R}\rangle_{\rho_{\ast}} \label{EID-07}%
\end{equation}
where $\rho_{\ast}$ is such that $\rho_{S\ast}$ results from the projection of
$\rho_{\ast}$ onto $\mathcal{O}_{R}$. Therefore, for any observable
$O_{R}\subset$ $\mathcal{O}_{R}$,
\begin{equation}
\lim_{t\rightarrow\infty}\langle O_{R}\rangle_{\rho(t)}=\langle O_{R}%
\rangle_{\rho_{\ast}} \label{EID-08}%
\end{equation}
This result can also be expressed as a weak limit
\begin{equation}
W-\lim_{t\rightarrow\infty}\rho(t)=\rho_{\ast} \label{EID-09}%
\end{equation}

If the just obtained eq. (\ref{EID-08}) is compared with eq. (\ref{INT-01}),
it turns out to be clear that the EID approach can also be formulated from the
viewpoint of the closed composite system $U$ and, from this perspective, it
can be explained in the context of the general framework introduced in the
Introduction. In other words, the split of the closed system into a proper
open system and an environment is just a way of selecting the relevant
observables of the closed system.

The limit of eqs. (\ref{EID-06}) and (\ref{EID-07}) can be computed by means
of two different strategies:

\begin{description}
\item[.] By solving the unitary evolution equation for $\rho(t)$, computing
$Tr\,\left(  \rho(t)\,O_{R}\right)  $ and finding the limit.

\item[.] By solving the non-unitary evolution equation for $\rho_{S}%
(t)=Tr_{E}\left(  \rho(t)\right)  $, computing $Tr\left(  \rho_{S}%
(t)\,O_{S}\right)  $ and finding the limit.
\end{description}

Of course, the two strategies give the same result: although the second is the
usual method in the EID literature, the first may lead to a simpler solution,
as in the model developed in the next section.

\section{The spin-bath model from the general framework}

The spin-bath model is a very simple model that has been exactly solved in
previous papers (see \cite{Zurek-1982}). Here we will study it from the
general framework applied to the EID approach, as presented in the previous
section. This task will allow us to compare the method traditionally used in
the literature for solving the model with the method based on our lemma.

Let us consider a closed system $U=P\cup P_{1}\cup\ldots\cup P_{N}=P\cup
(\cup_{i=1}^{N}P_{i})$, where (i) $P$ is a spin-1/2 particle represented in
the Hilbert space $\mathcal{H}_{P}$, and (ii) each $P_{i}$ is a spin-1/2
particle represented in its Hilbert space $\mathcal{H}_{i}$. The Hilbert space
of the composite system $U$\ is, then,
\begin{equation}
\mathcal{H}=\mathcal{H}_{P}\otimes\left(  \bigotimes\limits_{i=1}%
^{N}\mathcal{H}_{i}\right)  \label{3.0}%
\end{equation}
In the particle $P$, the two eigenstates of the spin operator
$S_{P,\overrightarrow{v}}$\ in direction $\overrightarrow{v}$ are $\left\vert
\Uparrow\right\rangle ,\left\vert \Downarrow\right\rangle $:%
\begin{equation}
S_{P,\overrightarrow{v}}\left\vert \Uparrow\right\rangle =\frac{1}%
{2}\left\vert \Uparrow\right\rangle \ \ \ \ \ \ \ \ S_{P,\overrightarrow{v}%
}\left\vert \Downarrow\right\rangle =-\frac{1}{2}\left\vert \Downarrow
\right\rangle \label{3.1}%
\end{equation}
In each particle $P_{i}$, the two eigenstates of the corresponding spin
operator $S_{i,\overrightarrow{v}}$\ in direction $\overrightarrow{v}$ are
$\left\vert \uparrow_{i}\right\rangle ,\left\vert \downarrow_{i}\right\rangle
$:%
\begin{equation}
S_{i,\overrightarrow{v}}\left\vert \uparrow_{i}\right\rangle =\frac{1}%
{2}\left\vert \uparrow_{i}\right\rangle \text{ \ \ \ \ \ \ \ }%
S_{i,\overrightarrow{v}}\left\vert \downarrow_{i}\right\rangle =-\frac{1}%
{2}\left\vert \downarrow_{i}\right\rangle \label{3.2}%
\end{equation}
Therefore, a pure initial state of $U$ reads%
\begin{equation}
|\psi_{0}\rangle=(a\left\vert \Uparrow\right\rangle +b\left\vert
\Downarrow\right\rangle )\otimes\left(  \bigotimes_{i=1}^{N}(\alpha
_{i}|\uparrow_{i}\rangle+\beta_{i}|\downarrow_{i}\rangle)\right)  \label{3.3}%
\end{equation}
where $\left\vert a\right\vert ^{2}+\left\vert b\right\vert ^{2}=1$ and
$\left\vert \alpha_{i}\right\vert ^{2}+\left\vert \beta_{i}\right\vert ^{2}=1$
(for a generalization with $M$ spins $\left\{  \Uparrow,\Downarrow\right\}  $
and N spins $\left\{  \uparrow,\downarrow\right\}  $, see \cite{JPA-Sebi}). If
the self-Hamiltonians $H_{P}$ of $P$ and $H_{i}$ of $P_{i}$ are taken to be
zero, and there is no interaction among the $P_{i}$, then the total
Hamiltonian $H$ of the composite system $U$ is given by the interaction
between the particle $P$ and each particle $P_{i}$ (see \cite{Zurek-1982},
\cite{Max}):%
\begin{equation}
H=\frac{1}{2}\left(  \left\vert \Uparrow\right\rangle \left\langle
\Uparrow\right\vert -\left\vert \Downarrow\right\rangle \left\langle
\Downarrow\right\vert \right)  \otimes\sum_{i=1}^{N}\left[  g_{i}\left(
\left\vert \uparrow_{i}\right\rangle \left\langle \uparrow_{i}\right\vert
-\left\vert \downarrow_{i}\right\rangle \left\langle \downarrow_{i}\right\vert
\right)  \otimes\left(  \bigotimes_{j\neq i}^{N}\mathbb{I}_{j}\right)
\right]  \label{3.4}%
\end{equation}
where $\mathbb{I}_{j}=\left\vert \uparrow_{j}\right\rangle \left\langle
\uparrow_{j}\right\vert +\left\vert \downarrow_{j}\right\rangle \left\langle
\downarrow_{j}\right\vert $ is the identity operator on the subspace
$\mathcal{H}_{j}$. Under the action of $H$, the state $|\psi_{0}\rangle$
evolves into%
\begin{equation}
\left\vert \psi(t)\right\rangle =a\left\vert \Uparrow\right\rangle
|\mathcal{E}_{\Uparrow}(t)\rangle+b\left\vert \Downarrow\right\rangle
|\mathcal{E}_{\Downarrow}(t)\rangle\label{3.5}%
\end{equation}
where%
\begin{equation}
\left\vert \mathcal{E}_{\Uparrow}(t)\right\rangle =\left\vert \mathcal{E}%
_{\Downarrow}(-t)\right\rangle =\bigotimes_{i=1}^{N}\left(  \alpha
_{i}\,e^{-ig_{i}t/2}\,\left\vert \uparrow_{i}\right\rangle +\beta
_{i}\,e^{ig_{i}t/2}\,\left\vert \downarrow_{i}\right\rangle \right)
\label{3.6}%
\end{equation}

\subsection{Computing the expectation values}

The space $\mathcal{O}$ of the observables of the composite system $U$ can be
obtained as $\mathcal{O}=\mathcal{O}_{P}\otimes(\otimes_{i=1}^{N}%
\mathcal{O}_{i})$, where $\mathcal{O}_{P}$ is the space of the observables of
the particle $P$ and $\mathcal{O}_{i}$ is the space of the observables of the
particle $P_{i}$. Then, an observable $O\in\mathcal{O}=\mathcal{H}%
\otimes\mathcal{H}$ can be expressed as%
\begin{equation}
O=O_{P}\otimes(\bigotimes_{i=1}^{N}O_{i}) \label{3.7.1}%
\end{equation}
where%
\begin{align}
O_{P}  &  =s_{\Uparrow\Uparrow}\left\vert \Uparrow\right\rangle \left\langle
\Uparrow\right\vert +s_{\Uparrow\Downarrow}\left\vert \Uparrow\right\rangle
\left\langle \Downarrow\right\vert +s_{\Downarrow\Uparrow}\left\vert
\Downarrow\right\rangle \left\langle \Uparrow\right\vert +s_{\Downarrow
\Downarrow}\left\vert \Downarrow\right\rangle \left\langle \Downarrow
\right\vert \ \in\mathcal{O}_{P}\label{3.7.2}\\
O_{i}  &  =\epsilon_{\uparrow\uparrow}^{(i)}|\uparrow_{i}\rangle
\langle\uparrow_{i}|+\epsilon_{\downarrow\downarrow}^{(i)}|\downarrow
_{i}\rangle\langle\downarrow_{i}|+\epsilon_{\downarrow\uparrow}^{(i)}%
|\downarrow_{i}\rangle\langle\uparrow_{i}|+\epsilon_{\uparrow\downarrow}%
^{(i)}|\uparrow_{i}\rangle\langle\downarrow_{i}|\ \in\mathcal{O}_{i}
\label{3.7.3}%
\end{align}
Since the operators $O_{P}$ and $O_{i}$ are Hermitian, the diagonal components
$s_{\Uparrow\Uparrow}$, $s_{\Downarrow\Downarrow}$, $\epsilon_{\uparrow
\uparrow}^{(i)}$, $\epsilon_{\downarrow\downarrow}^{(i)}$ are real numbers,
and the off-diagonal components are complex numbers satisfying $s_{\Uparrow
\Downarrow}=s_{\Downarrow\Uparrow}^{\ast}$, $\epsilon_{\uparrow\downarrow
}^{(i)}=\epsilon_{\downarrow\uparrow}^{(i)\ast}$. Then, the expectation value
of the observable $O$ in the state $\left\vert \psi(t)\right\rangle $ of eq.
(\ref{3.5}) can be computed as%
\begin{equation}
\langle O\rangle_{\psi(t)}=(|a|^{2}s_{\Uparrow\Uparrow}+|b|^{2}s_{\Downarrow
\Downarrow})\,\Gamma_{0}(t)+2\operatorname{Re}\,[ab^{\ast}\,s_{\Downarrow
\Uparrow}\,\Gamma_{1}(t)] \label{3.8}%
\end{equation}
where (see \cite{Max})%
\begin{align}
\Gamma_{0}(t)  &  =\prod_{i=1}^{N}\left[  |\alpha_{i}|^{2}\epsilon
_{\uparrow\uparrow}^{(i)}+|\beta_{i}|^{2}\epsilon_{\downarrow\downarrow}%
^{(i)}+2\operatorname{Re}(\alpha_{i}{}\,\beta_{i}^{\ast}\epsilon
_{\downarrow\uparrow}^{(i)}e^{ig_{i}t})\right] \label{3.9}\\
\Gamma_{1}(t)  &  =\prod_{i=1}^{N}\left[  |\alpha_{i}|^{2}\epsilon
_{\uparrow\uparrow}^{(i)}e^{ig_{i}t}+|\beta_{i}|^{2}\epsilon_{\downarrow
\downarrow}^{(i)}e^{-ig_{i}t}+2\operatorname{Re}(\alpha_{i}{}\,\beta_{i}%
^{\ast}\epsilon_{\downarrow\uparrow}^{(i)})\right]  \label{3.10}%
\end{align}

\subsection{Selecting the relevant observables}

In the typical situation studied by the EID approach, the open system $S$ is
the particle $P$, and the remaining particles $P_{i}$ play the role of the
environment $E$: $S=P$ and $E=\cup_{i=1}^{N}P_{i}$. Therefore, the relevant
observables $O_{R}$ of the closed system $U$ are those corresponding to the
particle $P$, and they are obtained from eqs. (\ref{3.7.1}), (\ref{3.7.2}) and
(\ref{3.7.3}), by making $\epsilon_{\uparrow\uparrow}^{(i)}=\epsilon
_{\downarrow\downarrow}^{(i)}=1$ and $\epsilon_{\uparrow\downarrow}^{(i)}=0$:%
\begin{equation}
O_{R}=O_{S}\otimes\mathbb{I}_{E}=\left(  \sum_{s,s^{\prime}=\Uparrow
,\Downarrow}s_{ss^{\prime}}|s\rangle\langle s^{\prime}|\right)  \otimes\left(
\bigotimes_{i=1}^{N}\mathbb{I}_{i}\right)  \label{3.11}%
\end{equation}
The expectation value of these observables in the state $\left\vert
\psi(t)\right\rangle $ of eq. (\ref{3.5}) is given by%
\begin{equation}
\langle O_{R}\rangle_{\psi(t)}=|a|^{2}\,s_{\Uparrow\Uparrow}+|b|^{2}%
\,s_{\Downarrow\Downarrow}+2\operatorname{Re}[ab^{\ast}\,s_{\Downarrow
\Uparrow}\,r(t)] \label{3.12}%
\end{equation}
where
\begin{equation}
r(t)=\langle\mathcal{E}_{\Downarrow}(t)\rangle|\mathcal{E}_{\Uparrow
}(t)\rangle=\prod_{i=1}^{N}\left(  |\alpha_{i}|^{2}\,e^{-ig_{i}t}+|\beta
_{i}|^{2}\,e^{ig_{i}t}\right)  \label{3.13}%
\end{equation}
This means that, in eq. (\ref{3.8}), $\Gamma_{0}(t)=1$ and $\Gamma
_{1}(t)=r(t)$.

\section{Simulation versus prediction}

\subsection{Simulation: the usual method}

In order to know the time-behavior of the expectation value of eq.
(\ref{3.12}), the time-behavior of $r(t)$ has to be computed. From eq.
(\ref{3.12}) $|r(t)|^{2}$ results
\begin{equation}
|r(t)|^{2}=\prod_{i=1}^{N}(|\alpha_{i}|^{4}+|\beta_{i}|^{4}+2|\alpha_{i}%
|^{2}|\beta_{i}|^{2}\cos2g_{i}t) \label{3.14}%
\end{equation}
If $\left\vert \alpha_{i}\right\vert ^{2}$ and $\left\vert \beta
_{i}\right\vert ^{2}$ are taken as random numbers in the closed interval
$\left[  0,1\right]  $, such that $|\alpha_{i}|^{2}+|\beta_{i}|^{2}=1$, then
\begin{align}
\max_{t}(|\alpha_{i}|^{4}+|\beta_{i}|^{4}+2|\alpha_{i}|^{2}|\beta_{i}|^{2}%
\cos2g_{i}t)  &  =\left(  \left(  |\alpha_{i}|^{2}+|\beta_{i}|^{2}\right)
^{2}\right)  =1\nonumber\\
\min_{t}\left(  \left\vert \alpha_{i}\right\vert ^{4}+\left\vert \beta
_{i}\right\vert ^{4}+2\left\vert \alpha_{i}\right\vert ^{2}\left\vert
\beta_{i}\right\vert ^{2}\cos\left(  2g_{i}t\right)  \right)   &  =\left(
\left(  |\alpha_{i}|^{2}-|\beta_{i}|^{2}\right)  ^{2}\right)  =\left(
2\left\vert \alpha_{i}\right\vert ^{2}-1\right)  ^{2} \label{3.15}%
\end{align}
Therefore, $(|\alpha_{i}|^{4}+|\beta_{i}|^{4}+2|\alpha_{i}|^{2}|\beta_{i}%
|^{2}\cos2g_{i}t)$ is a random number which, if $t\neq0$, fluctuates between
$1$ and $\left(  2\left\vert \alpha_{i}\right\vert ^{2}-1\right)  ^{2}$.

In order to obtain the limit of $r(t)$ for $t\rightarrow\infty$, different
numerical simulations are performed and presented in the literature, where the
aleatory numbers $\left\vert \alpha_{i}\right\vert ^{2}$ and $\left\vert
\beta_{i}\right\vert ^{2}$ are obtained from a generator of aleatory numbers:
the generator fixed the value of $\left\vert \alpha_{i}\right\vert ^{2}$, and
the $\left\vert \beta_{i}\right\vert ^{2}$ is computed as $\left\vert
\beta_{i}\right\vert ^{2}=1-\left\vert \alpha_{i}\right\vert ^{2}$. The value
of the $g_{i}$ and the time interval $\left[  0,t_{0}\right]  $ for the
computations is usually stipulated. In general, the model is studied and the
conclusions about decoherence are drawn by means of this kind of numerical
simulations (\cite{Zurek-1982}, \cite{PLA}, \cite{Max}, \cite{Letter-1}).

\subsection{Prediction: using the lemma}

The first step for applying the lemma is to express the expectation values in
the energy eigenbasis. In this model the energy eigenbasis coincides with the
eigenbasis of the spin in direction $z$, which is obtained in terms of the
tensorial product of the eigenstates of the spin in direction $z$ for all the
particles. In other words, the eigenvectors of $H$, which form a basis of
$\mathcal{H}$, are%
\begin{align}
&  \left\vert \Uparrow\right\rangle \left\vert \uparrow_{1}\right\rangle
...\left\vert \uparrow_{k}\right\rangle ...\left\vert \uparrow_{N-1}%
\right\rangle \left\vert \uparrow_{N}\right\rangle \nonumber\\
&  \left\vert \Uparrow\right\rangle \left\vert \uparrow_{1}\right\rangle
...\left\vert \uparrow_{k}\right\rangle ...\left\vert \uparrow_{N-1}%
\right\rangle \left\vert \downarrow_{N}\right\rangle \nonumber\\
&  ...\nonumber\\
&  \left\vert \Downarrow\right\rangle \left\vert \downarrow_{1}\right\rangle
...\left\vert \downarrow_{k}\right\rangle ...\left\vert \downarrow
_{N-1}\right\rangle \left\vert \downarrow_{N}\right\rangle \label{3.17}%
\end{align}
Then, it is easy to see that in eq. (\ref{3.4}) $H$ is written in its diagonal
form, and that the expectation value of eq. (\ref{3.12}) is expressed in the
energy eigenbasis.

In order to simplify the expressions, we will introduce a particular
arrangement into the set of those eigenvectors by calling them $\left\vert
\mathcal{A}_{i}\right\rangle $:\ the set $\left\{  \left\vert \mathcal{A}%
_{i}\right\rangle \right\}  $ is an eigenbasis of $H$ with $2^{N+1}$ elements.
The $\left\vert \mathcal{A}_{i}\right\rangle $ will be ordered in terms of
their eigenvalues, which depend on the number of particles of $E$ having spin
$\left\vert \downarrow\right\rangle $ in any state. Then, we have:

\begin{itemize}
\item Two states where all the particles of $E$ have spin $\left\vert
\uparrow\right\rangle $:%
\begin{align}
\left\vert \mathcal{A}_{1}\right\rangle  &  =\left\vert \Uparrow
,\uparrow,...,\uparrow,\uparrow\right\rangle \Longrightarrow H\left\vert
\mathcal{A}_{1}\right\rangle =\frac{1}{2}\left(
{\textstyle\sum_{i=1}^{N}}
g_{i}\right)  \left\vert \mathcal{A}_{1}\right\rangle \nonumber\\
\left\vert \mathcal{A}_{-1}\right\rangle  &  =\left\vert \Downarrow
,\uparrow,...,\uparrow,\uparrow\right\rangle \Longrightarrow H\left\vert
\mathcal{A}_{-1}\right\rangle =-\frac{1}{2}\left(
{\textstyle\sum_{i=1}^{N}}
g_{i}\right)  \left\vert \mathcal{A}_{-1}\right\rangle \label{4.9}%
\end{align}

\item $2N$ states where only one particle of $E$ has spin $\left\vert
\downarrow\right\rangle $:%
\begin{align}
\left\vert \mathcal{A}_{j}\right\rangle  &  =\left\vert \Uparrow
,\uparrow,...,\uparrow,\downarrow,\uparrow,...,\uparrow,\uparrow\right\rangle
\Longrightarrow H\left\vert \mathcal{A}_{j}\right\rangle =\frac{1}{2}\left(
{\textstyle\sum_{i=1}^{N}}
g_{i}-g_{k}\right)  \left\vert \mathcal{A}_{j}\right\rangle \nonumber\\
\left\vert \mathcal{A}_{-j}\right\rangle  &  =\left\vert \Downarrow
,\uparrow,...,\uparrow,\downarrow,\uparrow,...,\uparrow,\uparrow\right\rangle
\Longrightarrow H\left\vert \mathcal{A}_{-j}\right\rangle =-\frac{1}{2}\left(
%
{\textstyle\sum_{i=1}^{N}}
g_{i}-g_{k}\right)  \left\vert \mathcal{A}_{-j}\right\rangle \nonumber\\
\text{with \ \ }j  &  =2,3,...,N+1\text{ and }k=1,2,...,N \label{4.10}%
\end{align}

\item $\left(  N-1\right)  N$ states where two particles have spin $\left\vert
\downarrow\right\rangle $:%
\begin{align}
\left\vert \mathcal{A}_{j}\right\rangle  &  =\left\vert \Uparrow
,\uparrow,,...,\uparrow,\downarrow,\uparrow,,...,\uparrow,\downarrow
,\uparrow,...,\uparrow,\uparrow\right\rangle \Longrightarrow H\left\vert
\mathcal{A}_{j}\right\rangle =\frac{1}{2}\left(
{\textstyle\sum_{i=1}^{N}}
g_{i}-g_{k}-g_{l}\right)  \left\vert \mathcal{A}_{j}\right\rangle \nonumber\\
\left\vert \mathcal{A}_{-j}\right\rangle  &  =\left\vert \Uparrow
,\uparrow,,...,\uparrow,\downarrow,\uparrow,,...,\uparrow,\downarrow
,\uparrow,...,\uparrow,\uparrow\right\rangle \Longrightarrow H\left\vert
\mathcal{A}_{-j}\right\rangle =-\frac{1}{2}\left(
{\textstyle\sum_{i=1}^{N}}
g_{i}-g_{k}-g_{l}\right)  \left\vert \mathcal{A}_{-j}\right\rangle \nonumber\\
\text{with \ \ }j  &  =N+2,N+3,...,N+1+\frac{\left(  N-1\right)  N}{2}\text{
and }k,l=1,2,...,N \label{4.11}%
\end{align}

\item For the remaining particles with more spins $\left\vert \downarrow
\right\rangle $, the procedure is analogous.\medskip
\end{itemize}

Let us consider the two extreme cases. If the coupling coefficients $g_{i}%
$\ are random numbers, in principle all eigenvectors are different. If the
coupling coefficients are $g_{i}=g$, then there are%
\begin{align}
&  2\text{ eigenvectors with eigenvalue }\frac{N}{2}g\nonumber\\
&  2N\text{ eigenvectors with eigenvalue }\frac{N-2}{2}g\nonumber\\
&  \vdots\nonumber\\
&  2\frac{N!}{(N-l)!l!}\text{ eigenvectors with eigenvalue }\frac{N-2l}{2}g
\label{4.12}%
\end{align}
with $l=0,1,...N$. In this case, $H$ is degenerate: it has $2^{N+1}$
eigenvectors but only $2N$ different eigenvalues. In both cases, random
$g_{i}$ or equal $g_{i}$,\ the number of different possible energies is large
when $N$ is large enough.

On the other hand, eq. (\ref{3.12}) can be written as%
\begin{equation}
\langle O_{R}\rangle_{\psi(t)}=\sum_{i}\rho_{i}O_{i}\,+\sum_{\nu}\rho_{\nu
}^{\ast}O_{\nu}\,e^{i\omega_{\nu}t} \label{4.13}%
\end{equation}

where%
\begin{equation}
\sum_{i}\rho_{i}O_{i}=|a|^{2}\,s_{\Uparrow\Uparrow}+|b|^{2}\,s_{\Downarrow
\Downarrow}\label{4-14}%
\end{equation}%
\begin{equation}
\sum_{\nu}\rho_{\nu}^{\ast}O_{\nu}\,e^{i\omega_{\nu}t}=2\operatorname{Re}%
[ab^{\ast}\,s_{\Downarrow\Uparrow}\,\prod_{i=1}^{N}\left(  |\alpha_{i}%
|^{2}\,e^{-ig_{i}t}+|\beta_{i}|^{2}\,e^{ig_{i}t}\right)  ]\label{4-15}%
\end{equation}
The r.h.s. of this last expression includes a binomial product which can be
rewritten as a sum by means of the following strategy. First we define the
index $\nu$ that establishes the number of the term of the sum: since in eq.
(\ref{4-15}) there are $2^{N}$ terms, then $\nu=0,1,\cdots2^{N}-1$. Then we
define the number $p_{\nu,i}$ as the $i$ digit of the number $\nu$ written in
the binary system. Moreover, each term of the sum is a product of $N$
exponentials of the form $e^{-ig_{i}t}$, which can be grouped into a single
exponential $e^{i\omega_{\nu}t}$. The $\omega_{\nu}$ are all possible
additions and subtractions between the coefficients $g_{i}$; so, a generic
$\omega_{\nu}$ can be computed as
\begin{equation}
\omega_{\nu}=\left(
{\textstyle\sum_{i=1}^{N}}
\left(  -1\right)  ^{p_{\nu,i}}g_{i}\right)  \label{4.16}%
\end{equation}
Precisely
\begin{align}
\nu &  =0=0\cdots000_{b}\longrightarrow\omega_{0}=\left(
{\textstyle\sum_{i=1}^{N}}
g_{i}\right)  \nonumber\\
\nu &  =1=0\cdots001_{b}\longrightarrow\omega_{1}=\left(
{\textstyle\sum_{i=1}^{N-1}}
g_{i}-g_{N}\right)  \nonumber\\
\nu &  =2=0\cdots010_{b}\longrightarrow\omega_{2}=\left(
{\textstyle\sum_{i=1}^{N-2}}
g_{i}-g_{N-1}+g_{N}\right)  \nonumber\\
\nu &  =3=0\cdots011_{b}\longrightarrow,\omega_{3}=\left(
{\textstyle\sum_{i=1}^{N-2}}
g_{i}-g_{N-1}-g_{N}\right)  \nonumber\\
\nu &  =4=0\cdots100_{b}\longrightarrow\omega_{4}=\left(
{\textstyle\sum_{i=1}^{N-2}}
g_{i}-g_{N-2}+g_{N-1}+g_{N}\right)  \nonumber\\
&  \vdots\nonumber\\
\nu &  =2^{N}-1=1\cdots1_{b}\longrightarrow\omega_{2^{N}}=\left(  -%
{\textstyle\sum_{i=1}^{N}}
g_{i}\right)  \label{4.17}%
\end{align}
On the basis of this strategy, we can define the discrete function
$f_{d}(\omega_{\nu})$\ as
\begin{equation}
f_{d}(\omega_{\nu})=\prod_{k=1}^{N}|\gamma_{\nu,k}|^{2}\label{4.18}%
\end{equation}
where $\gamma_{\nu,k}=\left(  \alpha_{k}-\beta_{k}\right)  p_{\nu,k}+\beta
_{k}$, which is equal to $\alpha_{k}$ if $p_{\nu,k}=1$\ and is equal to
$\beta_{k}$\ if $p_{\nu,k}=0$. Then, the binomial product of eq. (\ref{4-15})
results
\begin{equation}
\prod_{i=1}^{N}\left(  |\alpha_{i}|^{2}\,e^{-ig_{i}t}+|\beta_{i}%
|^{2}\,e^{ig_{i}t}\right)  =\sum_{\nu=0}^{2^{N}-1}f_{d}(\omega_{\nu
})e^{-i\omega_{\nu}t}\label{4.19}%
\end{equation}
then,%
\begin{equation}
\langle O_{R}\rangle_{\psi(t)}=\sum_{i}\rho_{i}O_{i}\,+2\operatorname{Re}%
[ab^{\ast}\,s_{\Downarrow\Uparrow}\,\sum_{\nu=0}^{2^{N}-1}f_{d}(\omega_{\nu
})\,e^{-i\omega_{\nu}t}]\label{4.20}%
\end{equation}
In order to apply our lemma to eq. (\ref{4.20}), it is necessary that
$f_{d}(\omega_{\nu})\in\mathcal{L}_{1}$. On the one hand, since $\left\{
\omega_{\nu}\right\}  $\ has $2^{N}$\ elements, then for $N\gg1$ the set
$\left\{  \omega_{\nu}\right\}  $ is quasi-continuous of class 1. On the other
hand, since $f_{d}(\omega_{\nu})$\ is defined in eq. (\ref{4.18}), where
$0<|\gamma_{\nu,k}|^{2}<1$,\ then $f_{d}(\omega_{\nu})$\ is the product of $N$
numbers lower than $1$. Therefore, if $N\gg1$,
\begin{align}
|\gamma_{\nu,k}|^{2} &  <1\Rightarrow f_{d}(\omega_{\nu})=\prod_{k=1}%
^{N}|\gamma_{\nu,k}|^{2}\ll1\text{ if }N\gg1\quad\Rightarrow\quad
0<f_{d}(\omega_{\nu})<\varepsilon\ll1\nonumber\\
\qquad &  \Rightarrow\max_{\mu,\nu}\left(  \left\vert f_{d}(\omega_{\mu
})-f_{d}(\omega_{\nu})\right\vert \right)  \ll1\label{4.21}%
\end{align}
But this is precisely the condition for $f_{d}(\omega_{\nu})\in\mathcal{L}%
_{1}$. As a consequence, according to the lemma,
\begin{equation}
\sum_{\nu}\rho_{\nu}^{\ast}O_{\nu}\,e^{i\omega_{\nu}t}\rightarrow
0\quad\Longrightarrow\quad\langle O_{R}\rangle_{\psi(t)}\rightarrow\sum
_{i}\rho_{i}O_{i}\label{4.22}%
\end{equation}
and we can conclude, with no need of computer simulations, that the system decoheres.

\section{Conclusions}

In another paper of this issue, a discrete analogue of the Riemann-Lebesgue
theorem is presented and, on this basis, a lemma relevant for discrete models
is introduced: such a lemma provides a criterion for deciding whether or not
the system decoheres with no need of numerical simulations. In order to
present an example of how these results can be usefully exploited for the
study of decoherence in discrete models, in this paper we have applied that
lemma to the well-known spin-bath model, and we have shown that the conclusion
drawn from that application agrees with the results obtained by means of
computer simulations in the previous literature.

\section{Acknowledgments}

This research was partially supported by grants of the University of Buenos
Aires, the CONICET and the FONCYT of Argentina.


\begin{thebibliography}{99}                                                                                               %
\bibitem {CQG-General}M. Castagnino, S. Fortin, R. Laura and O. Lombardi,
\textit{Class. Quantum Grav.}, \textbf{25}, 154002, 2008.

\bibitem {JPA-Sebi}M. Castagnino, S. Fortin and O. Lombardi, \textit{Jour.
Phys. A: Math. and Theo}r., \textbf{43}, 065304, 2010.

\bibitem {PLA}M. Castagnino, S. Fortin and O. Lombardi, \textit{Mod. Physics
Lett.} A, \textbf{25}, 1431, 2010.

\bibitem {Letter-1}M. Castagnino, S. Fortin and O. Lombardi, \textit{Mod.
Phys. Lett. A}, \textbf{25}, 611, 2010.

\bibitem {Discreto}M. Castagnino and S. Fortin, \textquotedblleft Predicting
decoherence in discrete models\textquotedblright, \textit{Int. Jour. Theor.
Phys.}, this issue, 2010.

\bibitem {Zeh-1970}H. D. Zeh, \textit{Found. Phys.}, \textbf{1}, 69, 1970

\bibitem {Zeh-1973}H. D. Zeh, \textit{Found. Phys.}, \textbf{3}, 109, 1973.

\bibitem {Zurek-1982}W. H. Zurek, \textit{Phys. Rev. D}, \textbf{26}, 1862, 1982.

\bibitem {Zurek-1993}W. H. Zurek, \textit{Progr. Theor. Phys}., \textbf{89},
281, 1993.

\bibitem {Paz-Zurek}J. P. Paz and W. Z., \textquotedblleft Environment-induced
decoherence and the transition from quantum to classical\textquotedblright, in
Dieter Heiss (ed.), \textit{Lecture Notes in Physics, Vol. 587}, Springer,
Heidelberg-Berlin, 2002.

\bibitem {Zurek-2003}W. H. Zurek, \textit{Rev. Mod. Phys}., \textbf{75}, 715, 2003.

\bibitem {Max}M. Schl\"{o}sshauer, \textit{Phys. Rev. A}, \textbf{72}, 012109, 2005.


\end{thebibliography}
\end{document}